\documentclass[useAMS,referee,usenatbib]{biom}

\usepackage[colorlinks=true, urlcolor=blue, linkcolor=blue, citecolor=blue]{hyperref}
\usepackage{url}
\newcommand{\blind}{0}
\newcommand{\blindedinline}[1]{
    \if0\blind {#1}\fi
    \if1\blind {\texttt{<redacted>}}\fi
}
\usepackage{todonotes}

\usepackage{amsmath}
\usepackage{amssymb}
\usepackage{bbm}
\usepackage{mathabx}

\newcommand{\bx}{{\mathbf x}}
\newcommand{\bt}{{\mathbf t}}
\newcommand{\by}{{\mathbf y}}
\newcommand{\bu}{{\mathbf u}}
\newcommand{\br}{{\mathbf r}}
\newcommand{\bk}{{\mathbf k}}
\newcommand{\boldm}{{\mathbf m}}
\newcommand{\bb}{{\mathbf b}}

\newcommand{\bX}{{\mathbf X}}
\newcommand{\bD}{{\mathbf D}}
\newcommand{\bI}{{\mathbf I}}
\newcommand{\bB}{{\mathbf B}}
\newcommand{\bP}{{\mathbf P}}
\newcommand{\bK}{{\mathbf K}}
\newcommand{\bV}{{\mathbf V}}

\newcommand{\bbeta}{{\boldsymbol \beta}}

\newcommand{\btau}{{\boldsymbol \tau}}

\newcommand{\bone}{{\mathbf 1}}
\newcommand{\bzero}{{\mathbf 0}}

\newcommand{\bbR}{{\mathbb R}}
\newcommand{\bbE}{{\mathbb E}}
\newcommand{\ind}{{\mathbbm 1}}

\newcommand{\cN}{\mathcal N}
\newcommand{\cP}{\mathcal P}
\newcommand{\eps}{{\varepsilon}}

\DeclareMathOperator{\Var}{Var}
\DeclareMathOperator{\Cov}{Cov}
\DeclareMathOperator{\diag}{diag}

\DeclareMathOperator{\clr}{clr}

\usepackage{soul}
\newcommand{\annotated}{0} 
\definecolor{antiquefuchsia}{rgb}{0.57, 0.36, 0.51}
\newcommand{\add}[1]{\if0\annotated{#1}\fi\if1\annotated{\textcolor{teal}{#1}}\fi}
\newcommand{\update}[1]{\if0\annotated{#1}\fi\if1\annotated{\textcolor{antiquefuchsia}{#1}}\fi}
\newcommand{\drop}[1]{\if0\annotated{}\fi\if1\annotated{\textcolor{orange}{#1}}\fi}
\newcommand{\replace}[2]{\drop{#1} \add{#2}}

\title[Locally sparse varying coefficient mixed model]{Locally sparse varying coefficient mixed model 
  with application to longitudinal microbiome differential abundance}

\author{
Simon Fontaine$^{1,6,*}$\email{simfont@umich.edu}, 
Nisha J. D'Silva$^{2,3}$, 
Marcell Costa de Medeiros$^{2}$\\
\textbf{
Grace Y. Chen$^{4}$, 
Ji Zhu$^{1}$, 
and Gen Li$^{5}$
} 
 \\
$^{1}$Department of Statistics, University of Michigan, Ann Arbor, MI, United States \\
$^{2}$Department of Periodontics and Oral Medicine, University of Michigan, Ann Arbor, MI, United States \\
$^{3}$Department of Pathology, University of Michigan, Ann Arbor, MI, United States \\
$^{4}$Department of Internal Medicine, University of Michigan, Ann Arbor, MI, United States \\
$^{5}$Department of Biostatistics, University of Michigan, Ann Arbor, MI, United States \\
$^{6}$Department of Statistics, Pennsylvania State University, University Park, PA, United States
}

\begin{document}


\date{{\it Received August} 2024. {\it Revised May} 2025.  {\it
Accepted August} 2025.}



\pagerange{\pageref{firstpage}--\pageref{lastpage}} 
\volume{68}
\pubyear{2025}
\artmonth{May}


\doi{10.1111/tbd}


\label{firstpage}


\begin{abstract}
Differential abundance (DA) analysis in microbiome studies has recently been used to uncover a plethora of associations between microbial composition and various health conditions. 
While current approaches to DA typically apply only to cross-sectional data, many studies feature a longitudinal design to better understand the underlying microbial dynamics. 
To study DA in longitudinal microbial studies, we introduce a novel varying coefficient mixed-effects model with local sparsity. 
The proposed method can identify time intervals of significant group differences while accounting for temporal dependence. 
Specifically, we exploit a penalized kernel smoothing approach for parameter estimation and
include a random effect to account for serial correlation. 
In particular, our method operates effectively regardless of whether sampling times are shared across subjects, accommodating irregular sampling and missing observations. 
Simulation studies demonstrate the necessity of modeling dependence for precise estimation and support recovery. 
The application of our method to a longitudinal study of mice oral microbiome during cancer development revealed significant scientific insights that were otherwise not discernible through cross-sectional analyses.
An \texttt{R} implementation is available at \href{https://github.com/fontaine618/LSVCMM}{github.com/fontaine618/LSVCMM}.
\end{abstract}

%



\maketitle


%
\if1\annotated
Color code: \add{Added}, \drop{Removed}, \update{Updated} 
\fi
\section{Introduction}
\label{sec:introduction}


\drop{\subsection{Longitudinal Differential Analysis in Omics Data}}\label{subsec:introduction.da}

The modern science and healthcare sectors have seen profound advancements with the emergence of omics data, such as (meta-)genomics, transcriptomics, and proteomics, among others. In particular, many studies collect omics longitudinally, which can provide researchers with greater insight into the dynamical complexities underlying a myriad of biological processes. A natural statistical question that emerges from temporal omics is that of longitudinal differential analysis (LDA), where the goal is to identify biomarkers with differences between conditions over time. 
\drop{A naïve approach to LDA would be to perform separate cross-sectional differential analyzes at a collection of time points of interest. However, this approach cannot borrow strength over time as two neighboring time points would be treated independently. Additionally, when samples are collected repeatedly on a set of subject, cross-sectional analysis ignores serial correlation, which can greatly impair statistical properties. Finally, cross-sectional methods often struggle with irregularly sampled time points as the number of samples at a given time point might be small, thus requiring preprocessing. In recent years, multiple methods for LDA have been proposed to account for some of these challenges \citep{staicu_significance_2015, luo_informative_2017, metwally_robust_2022}, though they are limited to simple group comparisons, whereas more complicated designs, such as interactions or confounder adjustment, are often of interest.} 
Our motivating application involves the field of \textit{microbiomics}, where differential analysis usually takes the name of \textit{differential abundance analysis} (DAA), referring to the \textit{abundance} of various organisms in a system of interest. Specifically, we are interested in the microbial composition in some tissue and its temporal association with multiple conditions simultaneously. The microbiota is known to play a key role in a wide range of diseases and health conditions \citep{gomaa_human_2020, ogunrinola_human_2020}, and identifying taxa that differ in their abundance between conditions can lead to better diagnosis, prevention, and treatment. 
\drop{For the task of longitudinal DAA, few specialized methodologies have been proposed \citep{paulson_longitudinal_2017, shields-cutler_splinectomer_2018, metwally_metalonda_2018, jeganathan_block_2018} and none satisfactorily addresses the particularities of LDA while being flexible enough for general regression designs.  }

\drop{\subsection{Oral Cancer Development Mouse Study}}\label{subsec:introduction.cancer}

\drop{The tumor suppressor gene Dmbt1 (deleted in malignant brain tumors 1) plays a critical role in the progression of oral squamous cell carcinoma (SCC) \citep{singh_squamous_2021}. Dmbt1 is also present in human saliva, where it has an antimicrobial role \citep{reichhardt_salsa-dance_2017, ligtenberg_human_2001}. In a recent longitudinal study, \blindedinline{\cite{medeiros_salivary_2023}} observed that Dmbt1 is suppressed in saliva from patients with SCC prior to treatment and is upregulated after treatment. Consistent with these findings, mouse saliva showed a decrease in Dmbt1 expression after induction of SCC, suggesting that SCC down-regulates Dmbt1 expression. Furthermore, pre- and posttreatment changes in saliva Dmbt1 levels were associated with changes in some bacterial populations. In addition, there were differences before and after treatment in salivary bacteria in patients who responded or did not respond to chemoradiation treatment. }

\drop{To better understand the interaction between Dmbt1, microbial composition, and oral SCC
development, a mouse study was conducted \blindedinline{(Medeiros et al, manuscript in preparation)}. Seventy-six (76) mice were
bred with (\textit{wild type}, WT) and without (\textit{knockout}, KO) the Dmbt1 gene before being
inoculated with oral SCC. The saliva samples were then collected over time (0, 4, 8, 12, 16 and 22 weeks
after inoculation) and 16S sequencing was performed (16S rRNA, 97\% sequence similarity OTU binning). Finally, the histopathology of the mouse tongue was evaluated at week 22 where the mice were diagnosed with pre-cancer 
\textit{epithelial dysplasia} (ED) or 
\textit{carcinoma in situ} (CIS), or SCC.
The results strengthen the original findings of \cite{singh_squamous_2021}
as 17 of the 34 (50\%) knockout mice and only 6 of the 42 (14\%) wild type mice developed SCC by week 22, suggesting a
\textit{causal} link between Dmbt1 and cancer progression. }

\drop{A potential avenue of action of Dmbt1 on cancer progression, as suggested by the
findings in \blindedinline{\cite{medeiros_salivary_2023}}, is through the microbiota.
Therefore, one of the objectives of the study is to
investigate the longitudinal association of microbial composition
with the Dmbt1 genotype (WT vs. KO) and with diagnosis (dichotomized as precancer, ED / CIS, vs. cancer, SCC) as well as the
interaction between genotype and diagnosis. In particular, the identification of specific OTUs and weeks
with differential abundance between any of the subgroups can reveal how Dmbt1 and the microbial composition
influence cancer progression, potentially leading to better prediction of treatment response
and individualized treatments.
To this end, we consider the following \textit{varying coefficient model} (VCM) for the
\textit{centered log-ratio} (CLR) transformed abundance of each OTU:
\begin{align}
    \bbE\{\clr(y(w))\mid \bx\}
    &=
    \beta_0(w) 
    + \beta_{\text{KO}}(w)x_{\text{KO}}\nonumber
    + \beta_{\text{SCC}}(w)x_{\text{SCC}} \\
    &\qquad+ \beta_{\text{KO:SCC}}(w)x_{\text{KO}}x_{\text{SCC}}
    + \beta_{\text{F}}(w)x_{\text{F}},
    \label{eq:introduction.vcm}
\end{align}
where $w$ denotes the week and where $x_{\text{KO}}, x_{\text{SCC}}$ and $x_{\text{F}}$
are group indicators for genotype, diagnosis and sex, respectively.
In particular, we consider five varying coefficient terms: an intercept $\beta_0(\cdot)$
(corresponding to the reference group of WT, ED/CIS and male), a main effect of
genotype $\beta_{\text{KO}}(\cdot)$ (corresponding to the KO--WT difference), 
a main effect of diagnosis $\beta_{\text{SCC}}(\cdot)$ (corresponding to the SCC--ED/CIS difference), an 
interaction term between genotype and diagnosis $\beta_{\text{KO:SCC}}(\cdot)$ (1 if KO and SCC; 0 otherwise), and a main effect of sex $\beta_{\text{F}}(\cdot)$
(corresponding to the F--M difference). We adjusted for sex in the model as it is suspected to be associated with microbial composition, but it is of lesser interest. Figure~\ref{fig:introduction.otu}
shows an example OTU with differential abundance between genotypes at week 8 obtained from
our proposed \texttt{LSVCMM} method for the simpler model including the longitudinal effect
of genotype only. }


\drop{An important challenge that emerges from data collection is a significant amount of missing data. Indeed, not all saliva samples were collected and sequenced. Saliva samples were collected for only 65 mice and, of the $6\times 65=390$ potential samples, only 294 (75\%) were ultimately sequenced. 
Figure~\ref{fig:dmbt.missingness} shows the number of samples available per week and per subgroup, along the missingness patterns. About half (35/65) of the mice were sequenced at all six time
points, while most of the remaining (28/65) are missing weeks 4, 8 and 12. Additionally, 
missingness differs with conditions: specifically, mice with slower progression (ED/CIS)
are missing samples more frequently.}


\drop{\subsection{Proposed Methodology}}\label{subsec:introduction.methodology}

\update{We begin by describing the desirable properties of an LDA method.
First, the estimated differences need to be smooth over time, as there is generally no expectation
of discontinuous jumps. Additionally, smoothing may increase the power of the method as it allows us to borrow strength across neighboring time points.
Second, the longitudinal nature of the data implies serial correlation within subjects, and failure to account for it may hinder the statistical properties of the method. 
Third, accomodating irregular designs demands a methodology that does not
require all subjects to be measured at a common set of time points. Indeed, many studies have missing data, or samples may be collected irregularly in observational studies. 
Fourth, the methodology must be able not only to identify if there is a difference between conditions (\textit{global} differences), but it must also determine the time interval(s) where the difference occurs (\textit{local} differences). 
Fifth, the methodology must allow for more complicated regression designs than simple group comparisons: for example, our motivating application involves interactions and covariate adjustments.}

\add{\textit{Varying coefficient models} (VCMs) are natural candidates for LDA because their flexible semiparametric linear model structure can be customized for many purposes.} Still, we found no existing methodology that satisfies all the above requirements.
Many early work for regularized VCMs only allows global sparsity with the goal of selecting
variables, not time points \citep{wang_variable_2008, wang_shrinkage_2009, noh_sparse_2010, lee_local_2016, xue_variable_2012, daye_sparse_2012}. 
The idea of local sparsity was first introduced by \cite{wang_functional_2015}
and \cite{kong_domain_2015} using B-splines and kernel smoothing, respectively.
When using B-splines, local sparsity can be achieved by penalizing groups of
consecutive spline weights, and some type of overlapping group penalty must be used. 
In \cite{wang_functional_2015}, a group bridge penalty is used to induce local sparsity
in a single varying intercept; \cite{tu_estimation_2020} extended the approach to
more general VCMs. \cite{zhong_locally_2022} consider instead an application of the
functional SCAD penalty of \cite{lin_locally_2017} 
and extend the methodology beyond least squares and to asynchronous covariates.
When using kernel smoothing, local sparsity is simpler to achieve, as the evaluation of the varying coefficients can be directly penalized. In \cite{kong_domain_2015},
which considers a local linear approximation, a SCAD penalty is applied to a combination
of the degree 0 and degree 1 parameters of the local linear model.
However, none of the aforementioned methods adjusts for temporal dependency in their estimation procedure.
\cite{wang_functional_2022} consider a B-spline approach with a group bridge penalty
on the spline weights and adjust for within-subject dependency using a two-step estimator.
However, estimation of serial correlation requires all subjects to be sampled
at a common set of time points, which can be prohibitive.

\update{There exist a few methods performing longitudinal
differential abundance analysis through testing rather than through sparsity-inducing penalties.
A first set of methods builds on the concept of \textit{smoothing spline analysis of variance} (SS-ANOVA,\citealp{gu_smoothing_2013}) by proposing area ratio test statistics
to find intervals of group differences between conditions
\citep{paulson_longitudinal_2017, luo_informative_2017, metwally_metalonda_2018, metwally_robust_2022}.
Alternatively, \cite{shields-cutler_splinectomer_2018} propose a LOESS approach,
where permutations are used to identify differential intervals. Again, the previous methods
do not account for the dependency within the subject.
\cite{staicu_significance_2015} propose a global test for equality of means between groups,
accounting for longitudinal effects and based on Fourier expansions. 
Unfortunately, all these testing procedures are limited to group comparisons. }

We therefore propose a novel approach for local sparsity in VCMs, accounting for within-subject dependency,
called \texttt{LSVCMM} (for \textit{locally sparse varying coefficient mixed model}).
Specifically, we consider locally-constant kernel smoothing for a VCM
with parametric working covariance and obtain local and global sparsity through an adaptive
sparse group Lasso \citep{friedman_note_2010, simon_sparse-group_2013}. \drop{Estimation
alternates between mean parameter updates and covariance parameter updates. The mean parameter estimates
are taken as the solution to penalized score equations leading to proximal operators; the
variance parameter estimates optimize a Gaussian quasi-likelihood.} An extended Bayesian information
criterion is proposed to perform tuning parameter selection, and simultaneous confidence bands
are obtained through a bootstrap procedure. An \texttt{R} implementation is provided
in the \texttt{LSMCMM} package available at \blindedinline{\url{github.com/fontaine618/LSVCMM}}.
In Section~\ref{sec:simulations}, we conduct extensive simulation studies showing that \texttt{LSVCMM}
improves estimation accuracy and support recovery compared to methods lacking
longitudinal dependence adjustment or smoothness, or methods requiring imputation in irregular designs. 
\drop{Additionally, we compare \texttt{LSVCMM}
to \texttt{SPFDA} \citep{wang_functional_2022}, which requires imputation, and show significant improvements.
In Section~\ref{sec:real}, we apply \texttt{LSVCMM} to VCM \eqref{eq:introduction.vcm} discussed
in Section~\ref{subsec:introduction.cancer}. In particular, we identify five candidate OTUs for which their
temporal trajectories vary with both the Dmbt1 genotype and the SCC diagnosis at week 22.
These findings suggest pathways of action between the Dmbt1 protein and cancer development through the
microbial composition, which are found to be plausible based on related literature}
\add{In Section~\ref{sec:real}, we apply \texttt{LSVCMM} to a longitudinal differential analysis task that involves microbiome samples from mice with varying sex, genotype and disease status.}

\section{Methods}
\label{sec:methods}

\subsection{Setting \& Notation}\label{subsec:methods.setting}

We consider the following function-on-scalar regression problem. Let $i=1,\ldots, N$
denote the $N$ sampling units (e.g., subjects). Let $t_{in}$, $n=1,\ldots, N_i$, denote
the sampling times for subject $i$ and define $\bt_i=(t_{i1},\ldots, t_{iN_i})$; 
we do not assume any structure on the $\bt_{i}$'s 
across subjects. The observed response for subject $i$ at time $t_{in}$ is denoted
$y_{in}=y_i(t_{in})\in\bbR$ and we define 
$\by_i=y_i(\bt_i)=(y_{i1},\ldots , y_{in_i})\in\bbR^{n_i}$ as the vector
of responses for subject $i$. 
\drop{For each subject, we split covariates into two categories:
those associated with time-varying effects, $\bx_{in}=\bx_i(t_{in})\in\bbR^{p}$, and those with constant effects, $\bu_i\in\bbR^{p_u}$. }
In the present exposition, we assume the covariates $\bx_{in}=\bx_i(t_{in})\in\bbR^{p}$ to be constant through time, indicated by the absence of time index in 
$\bx_i\equiv\bx_i(\cdot)$, 
but our proposed model and implementation readily works for $\bx_{in}\in\bbR^{p}$ 
varying with time,
provided it is observed at the same time points as the responses of subject $i$. 
\drop{In particular, we do not allow \textit{asynchronous} covariates (see, e.g., 
\citealp{zhong_locally_2022}, for a related method). }


\subsection{Varying Coefficient Mixed Model}\label{subsec:methods.vcmm}

Our main goal is to study the relationship between covariates $\bx_i$ and the functional
response $y_{i}(\cdot)$,
while accounting for temporal dependence within subjects and other covariates.
In particular, we are interested in identifying \textit{if}, \textit{when} and \textit{how}
$y_{i}(\cdot)$ changes with each entry in $\bx_i$. 
To this end, we consider a \textit{varying coefficient mixed model}:
\begin{align}
   \bbE\{y_{in} \mid \theta_{i}(t_{in})\}
    & = 
    \bbeta(t_{in})^\top \bx_i
    +\theta_i(t_{in}) &
    \Var(y_{in} \mid \theta_{i}(t_{in})) &= \sigma^2
\end{align}
with (conditional) independence across $i$ and $n$, 
where $\bbeta(\cdot):\bbR\to\bbR^{p}$ is the vector-valued function of 
time-varying coefficients,
and where $\theta_i(\cdot)$ is a random process capturing the temporal dependence.
In particular, we assume $\bbE\{\theta_i(t)\}\equiv 0$ with covariance
kernel $\Cov(\theta_i(t), \theta_i(t'))=\sigma^2k_\theta(t, t')$ for some symmetric
positive definite kernel $k_\theta$. Define $\bK_\theta(\bt)$ as the (unscaled) covariance matrix
for a random process evaluated at the time points in $\bt$, that is, 
$[\bK_\theta(\bt)]_{nn'}=k_\theta(t_{n},t_{n'})$. Hence, marginally, the response vector $\by_i$ has mean $\boldm_i:=  \bbeta(\bt_i)^\top\bx_i$ and variance $\bV_i:=\sigma^2 \left(\bK_\theta(\bt_i) + \bI_{N_i}\right)$,
where $\bbeta(\bt_i)$ is the $p\times N_i$ matrix with columns $\bbeta(t_{in})$.
Given $S$ time points of interest $\bt=(t^{(1)}, \ldots, t^{(S)})$, we are interested
in the value of $\bbeta(\cdot)$ at each of those time points. For example, $\bt$
could consists of all observed time points or of a regular grid over the observed domain.
We define $\bB$ as the $p\times S$ matrix with entries $b^{(s)}_j=\bbeta_j(t^{(s)})$,
with rows $\bb_j=\beta_j(\bt)$ and with columns $\bb^{(s)}=\bbeta(t^{(s)})$. 

To obtain the most efficient estimator of the mean parameters, we need $\bV_i$ to
accurately capture the dependence structure in the residuals $\br_i = \by_i - \boldm_i$. For a regular design,
i.e, $\bt_i\equiv \bt$ for some $\bt$, and given $N$ sufficiently large, 
we could simply estimate
$\bV \equiv\bV_i \approx \frac{1}{N}\sum_{i=1}^{N} \br_i\br_i^\top$, perhaps with a prior smoothing
step \citep{wang_functional_2022}.
For irregular designs, more care is required as empirical covariances are not applicable.

As observed by \cite{fan_analysis_2007}, efficiency gains can be obtained even though the 
working covariance does not exactly match the true covariance. In particular, even a 
rough optimization of the working covariance can lead to near-optimal estimation
efficiency. We propose to specify a working parametric model whose covariance function
is determined by a few parameters. Some notable examples include the compound symmetry structure,
equivalent to a random intercept model, with covariance function
$
    k_\theta(t, s;r_\theta) = r_\theta,
$
and the AR(1) model, with covariance function
$
    k_\theta(t, s; r_\theta, \rho) = r_\theta \rho^{|t-s|},
$
where $r_\theta$ denotes the variance ratio with the noise variance $\sigma^2$ and 
where $\rho$ controls the long-range dependency. \drop{Given a working
covariance model, we denote by $\btau$ the set of all covariance parameters.}

\subsection{Local and Global Sparsity}\label{subsec:lsvcmm.sparsity}

A convenient feature of local regression is that the value of $\bbeta(\cdot)$ at a specific
time $t^{(s)}$ is directly parameterized by $\bb^{(s)}$. This is in contrast to spline 
basis expansions, where the value of $\bbeta(t^{(s)})$
is a linear combination of basis functions active at time $t^{(s)}$. Hence, to find 
$\beta_j(t^{(s)})=0$, we only need $b_j^{(s)}=0$, compared to requiring a consecutive
set of spline weights to be zero. This allows us to use a simple sparsity-inducing
penalty on the entries of $\bB$, 
in comparison to overlapping group penalties used in spline methods
\citep{wang_functional_2015, tu_estimation_2020, wang_functional_2022, zhong_locally_2022}.

We thus propose to encourage local sparsity by including a Lasso penalty \citep{tibshirani_regression_1996}
on $\bB$, namely, 
$\lambda\sum_{j=1}^{p}\sum_{s=1}^{S}\omega_j^{(s)}\vert b_j^{(s)}\vert$, where
$\omega_j^{(s)}$ are weights and where $\lambda\geqslant 0$ is the regularization parameter. 
\drop{For example, in group comparison setting, we would have two covariates:
an intercept $x_{i1}=1$ and a group membership $x_{i2}\in\{0, 1\}$. Then
we are only interested in encouraging zeros for the group difference effect 
$\bb_2$; in that case, the intercept $\bb_1$ would not be penalized, 
which can be achieved by setting $\omega_0^{(s)}\equiv0$.}
While the Lasso penalty really only encourages \textit{pointwise} sparsity, the combination
of smoothing with this penalization encourages \textit{local} sparsity, that is, sparsity across
consecutive time points. 

Furthermore, we may be interested in encouraging $\beta_j(\cdot)\equiv 0$ altogether
to identify if the $j$th covariate has any association with the response. This suggests to 
add a group lasso penalty on the whole vector $\bb_j$, 
leading to the sparse group Lasso penalty \citep{friedman_note_2010, simon_sparse-group_2013}:
\begin{align}
    \cP_{\lambda,\alpha}(\bB; \Omega)
    :=
    \lambda\sum_{j=1}^{p}\left[
    (1-\alpha)\sqrt{S}\omega_j\Vert\bb_j\Vert_2
    +\alpha\sum_{s=1}^{S}\omega_j^{(s)}\vert b_j^{(s)}\vert
    \right]\label{eq:lsvcmm.sparsity.penalty}
\end{align}
where $\alpha\in[0,1]$ is a tuning parameter balancing between encouraging global sparsity
($\alpha=0$) and local sparsity ($\alpha=1$), where $\omega_j$ are weights for
the group lasso penalty, and where $\Omega$ contains all weights.
\add{We refer to Section~S2.4 of the Supplementary Materials for experiments on the choice of $\alpha$.}

The Lasso and group Lasso penalty are famously known for their estimation bias
due to the shrinkage applied to non-zero values. To alleviate this issue, we propose to use
adaptive penalties \citep{zou_adaptive_2006, nardi_asymptotic_2008, poignard_asymptotic_2020}, where the weights are set to 
$\omega_j=\Vert\hat\bb_j^{\text{mle}}\Vert_2^{-\gamma}$ and
$\omega_j^{(s)}=\vert \hat b_j^{(s),\text{mle}}\vert^{-\gamma}$
for some $\gamma > 0$, where $\widehat{\bB}^{\text{mle}}$ contains the 
coefficients estimated without penalty.

\subsection{Estimation}\label{subsec:lsvcmm.estimation}

We alternate between mean parameter updates 
and variance parameter updates by holding 
the other fixed. We find that there is generally very little change beyond the first cycle
and that a single variance update is often sufficient (similar to the
two-step estimator of \citealp{wang_functional_2022}).

\paragraph{Kernel Smoothing}\label{subsubsec:lsvcmm.estimation.score}

Denote by $\bX_i$ the $n_i\times p$ matrix with rows $\bx_{in}$;
for fixed covariates, we simply have $\bX_i=\bone_{n_i}\bx_i^\top$. 
We define the estimating function for $\bbeta(t)$ by 
weighing the residuals using a kernel function $k_h(s)=k(s/h)/h$ depending 
on the distance from a time point of interest $t$, similarly to locally constant kernel smoothing
\citep{wang_efficient_2005}:
\begin{align}
    U_{\bbeta(t)}
    &:=
    -\sum_{i=1}^{N} \bX_i^\top \diag(\bk_i(t))\bV_i^{-1} \br_i,
\end{align}
where $\bk_i(t)=[k_h(t-t_{ij})]_{j=1}^{n_i}$, where $h>0$ is the kernel scale. 
We employ kernel smoothing rather than spline smoothing as inducing zeros
through penalization is simpler.

\drop{To motivate our GEE-based inference methodology, we start by investigating the score
equations for the mean parameters under a Gaussian model with likelihood 
\begin{align}
    \ell_i(\bbeta(\cdot))
    =
    -\frac{1}{2}\log\det(2\pi\bV_i)
    -\frac{1}{2}[\by_i-\boldm_i]^\top \bP_i[\by_i-\boldm_i],
\end{align}
with $\ell(\bbeta(\cdot)) = \sum_{i=1}^{N}\ell_i(\bbeta(\cdot)))$.
Consider computing the gradient with respect to $\bbeta(t)$ for some $t$.
Whenever $t_{ij}\neq t$, the mean $m_{ij}$ does not depend on $\bbeta(t)$,
so we find
\begin{align}
    \nabla_{\bbeta(t)}\ell(\bbeta(\cdot))
    &=
    -\sum_{i=1}^{N} \bX_i^\top \bD_i(t)\bP_i \br_i,
    \label{eq:score_pointwise_gradient}
\end{align}
where $\bD_i(t) = \diag(\ind[t=t_{ij}])$. Now, consider computing the gradient
with respect to $\bbeta$ by assuming that $\bbeta(\cdot)$ is a constant function
parameterized by $\bbeta$, i.e., $\bbeta(\cdot)\equiv \bbeta$. We find
\begin{align}
    \nabla_{\bbeta}\ell(\bbeta)
    &=
    -\sum_{i=1}^{N} \bX_i^\top \bP_i \br_i
    =
    -\sum_{i=1}^{N} \bX_i^\top \bI\bP_i \br_i.
    \label{eq:score_constant_gradient}
\end{align}
Looking at the difference between \eqref{eq:score_pointwise_gradient} and 
\eqref{eq:score_constant_gradient}, we see that the pointwise gradient
weighs the precision-adjusted residuals $\bP_i\br_i$ by $\bD_i(t)$, while the
constant gradient weighs them equally by $\bI$. To obtain a nonconstant smooth
estimate that borrows signal from neighboring time points, we utilize kernel smoothing,
which interpolates between the pointwise estimator and the constant estimator.
Specifically, we employ a locally-constant approximation around $t$ 
(a degree 0 local polynomial approximation), that is,
to estimate $\bbeta(t)$, we use the working model $\bbeta(t_{ij})\equiv \bbeta(t)$
and downweight observations with $t_{ij}$ away from $t$.
Of note, when $h\to 0$, $\diag(\bk_i(t))$ behaves as $\bD_i(t)$,
in which case we recover the pointwise estimator; 
when $h\to\infty$, $\diag(\bk_i(t))$ behaves as $\bI$,
in which case we recover the constant estimator. }

\paragraph{Proximal Updates}\label{subsubsec:lsvcmm.estimation.proximal}

The penalized estimating functions are given by adding the subgradients of the penalty
to the unpenalized functions, similarly to \cite{wang_penalized_2012} and
\cite{johnson_penalized_2008} where a linear approximation to the penalty is rather used:
\begin{align*}
    U_{\bb_j} + \partial_{\bb_j} P_{\lambda, \alpha}(\bB;\Omega),\qquad j=1,\ldots, p,
\end{align*}
where $U_{\bb_j}=(U_{b_j^{(1)}},\ldots, U_{b_j^{(S)}})\in\bbR^{S}$ is the estimating function
for covariate $j$, and where $\partial_{\bb_j}$ denotes the subgradient with respect to $\bb_j$.

Rather than considering a minorization-maximization scheme to solve the
penalized estimating equation 
$\bzero\in  U_{\bb_j} + \partial_{\bb_j} P_{\lambda, \alpha}(\bB;\Omega)$,
we utilize the convexity of the sparse group Lasso penalty
to our advantage and proceed to proximal gradient updates.
Specifically, we use the estimating function to perform a gradient step before
applying the corresponding proximal operator \citep{parikh_proximal_2014}.
We refer to Section~S1.1 of the Supplementary Material for details
on the proximal update.

\paragraph{Covariance Parameters Estimation}\label{subsubsec:lsvcmm.estimation.covariance}

To estimate the variance parameters $\btau$,
we maximize the profile likelihood under a Gaussian model (equivalently, a 
quasi-likelihood approach, \citealp{fan_semiparametric_2008})
while holding the residuals $\br_i$ fixed:
\begin{align*}
    \ell(\btau)
    :=
    -\frac{1}{2}
    \sum_{i=1}^{N}
    \log\det(2\pi\bV_i) + \br_i^\top \bV_i^{-1} \br_i,
\end{align*}
where $\bV_i$ implicitly depend on the variance parameters.
Updates for the compound symmetry
covariance can be found in Section~S1.2 of the Supplementary Material.

\subsection{Additional Details}
We refer to the Supplementary Material for additional information and 
practical guidelines about \texttt{LSVCMM}. In particular, we propose an
\textit{extended Bayesian information criterion} \citep[EBIC,][]{chen_extended_2008}
for selection of the regularization parameter $\lambda$ and the 
kernel scale $h$ (S1.3). Additionally, we utilize bootstrap sup-t simultaneous
confidence bands \citep{montiel_olea_simultaneous_2019} for uncertainty
quantification (S1.4), which also enable the computation of $p$-values that can be used for multiplicity corrections (S1.5).

\section{Simulation Studies}
\label{sec:simulations}

We consider two scenarios where the sampled time points $\bt_i$ vary between subjects.
In the first case, subjects are sampled on a common set of time points $\bt$, 
but not all time points are observed for each subject. This situation, inspired
from our real data application in Section~\ref{sec:real}, 
occurs in experimental studies with missing data.
In the second case, subjects are each sampled at different time points so
that none or few time points are shared across subjects. This situation more
commonly arises with observational studies, where researchers do not control
sampling.

Synthetic data is generated as follows. We consider the task of estimating
temporal group differences within $N=100$ subjects, half of which are assigned
to either group. \drop{For each subject $i$, sampled time points $\bt_i$ are generated
according to either scenario.} A random intercept $\theta_i(t)\equiv\theta_i$
is sampled from a normal distribution centered at 0 with variance $\sigma^2r_\theta$,
where $r_\theta$ denotes the variance ratio between noise and random effect. 
The mean function for each subject is computed as $\mu_i(t) = \beta_0(t) + \beta_1(t)x_i$,
where $x_i$ is the group indicator variable. Observations are finally
generated as $y_{ij} = \mu_i(t_{ij}) + \theta_i + \sigma \eps$ for $\eps\sim\cN(0,1)$.
Unless specifically varied in an experiment, default values for variance parameters
are $\sigma^2=1$ and $r_\theta=1$, leading to a correlation of 0.5 across time points. The true generating values for the time-varying effects 
$\beta_1(\cdot)$ will be defined in each sub-experiments such that 
it is smooth and non-zero only on part of the domain. 

For both scenarios, we conduct four sub-experiments. In the first experiment, 
we vary the signal strength by increasing the variance $\sigma^2$ while keeping
the signal fixed. In the second experiment, we investigate the effect of missing
data by varying the number of sampled time points per subject. In the third
experiment, we study the effect of the dependence by varying the 
variance ratio parameter $r_\theta$. \add{In the fourth experiment, we investigate the performance in relation to the sample size.}
We report the mean absolute estimation error (MAE) in estimating the functional group difference 
$\beta_1(\cdot)$
as well as the classification accuracy induced by the sparsity, both of which
over a pre-specified grid of time points. Additional performance metrics are included
in Section~S2.1 of the Supplementary Material.

Four methods are compared. Our proposed method, \texttt{LSVCMM}, is fitted
with a compound symmetry variance structure, using a Gaussian kernel with 
a fixed kernel scale ($h=0.2$)
and the regularization parameter is selected using the EBIC described in
Section~S1.3 of the Supplementary Materials. We also include \texttt{LSVCM}, which
is the same as \texttt{LSVCMM}, except that an independent variance structure is used.
\texttt{LSVCM} is closely related to the method of \cite{kong_domain_2015} where a 
local linear approximation and a SCAD penalty is rather utilized.
We further consider another simplification where the kernel smoothing is removed
(\texttt{ALasso}): this amounts to independent sparse estimation at each time point, 
though estimation of $\sigma^2$ and tuning parameter selection is done jointly.
Finally, we include \texttt{SPFDA} \citep{wang_functional_2022} which is a direct
competitor to \texttt{LSVCMM} as it includes smoothing, dependence and local sparsity, 
but it requires sampling times to be shared. To apply \texttt{SPFDA}, we impute
the missing samples using functional PCA \citep[fPCA,][]{goldsmith_corrected_2013}. 
\drop{In some cases, fPCA
fails due to a limited number of samples per subject; we proceed to mean
imputation instead for those instances. }
The bridge penalty parameter is set to $\alpha=0.5$ and the 
regularization parameter is selected using their propsed EBIC.

\subsection{Missing Data in Regular Design}

\begin{figure}
    \centering
    \includegraphics[width=\linewidth]{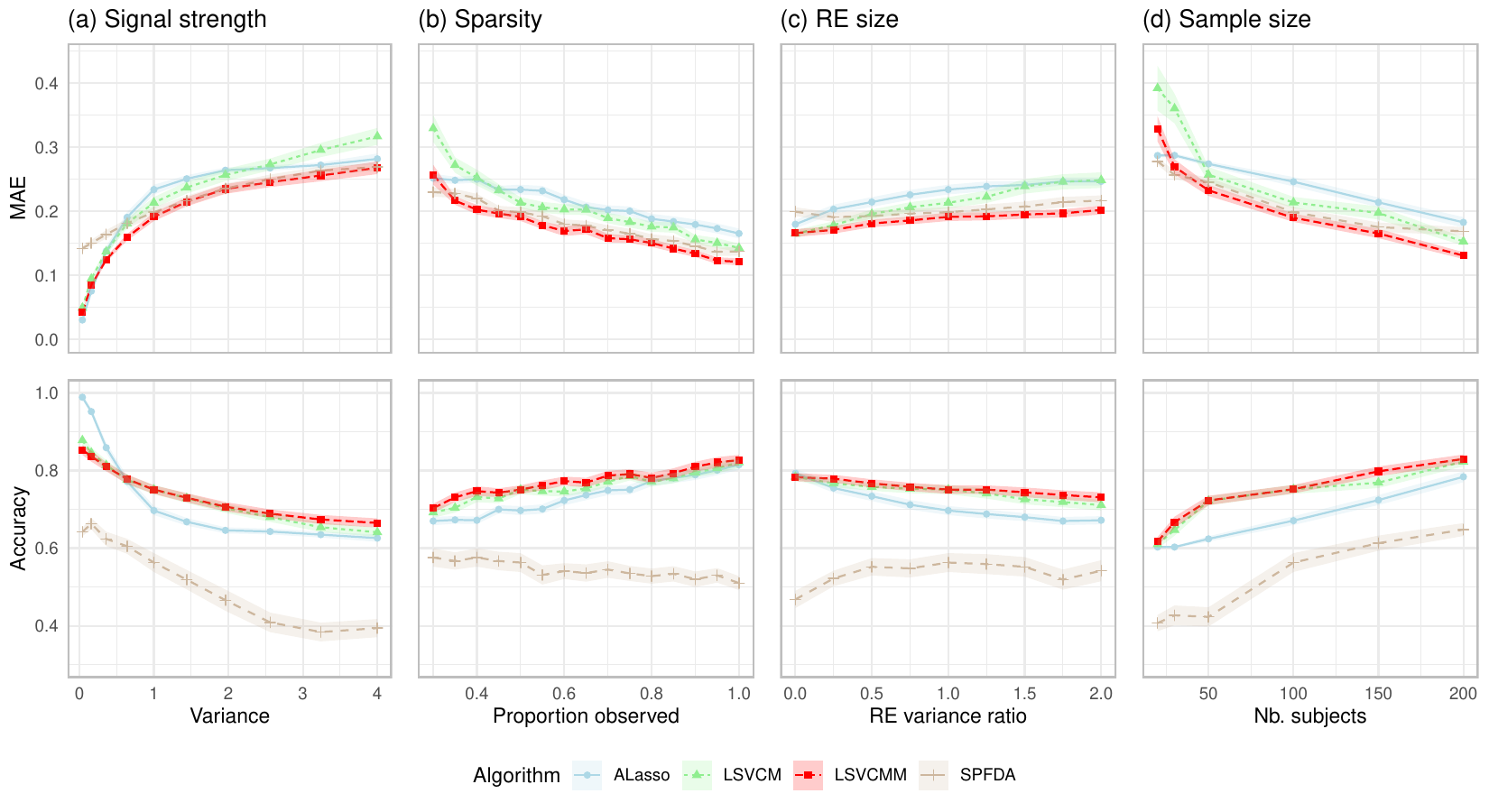}
    \caption{Evaluation metrics in the missing data scenario reported as the mean
    (line) and standard error (band) across 100 replications.}
    \label{fig:sim_block}
\end{figure}

In this scenario, we consider $10$ time points regularly-spaced on the unit interval.
To introduce missing data, we randomly select $71\%$ of the time points and
$71\%$ of the subjects and set the intersection as missing, leading to, on average,
$50\%$ of the $10\times 100$ samples to be missing. With this procedure, 
only $29\%$ of subjects will have 10 samples while the other $71\%$ will
only have around three. Symmetrically, three time points will be observed for
all 100 subjects, while the remaining seven will only be observed for 29 subjects.
The intercept function is chosen to be the zero function and the 
group difference function is set to be $\sin(2\pi(t-\tfrac 14))\vee 0$ so that
only the middle four time points are non-zero. \texttt{SPFDA} is fitted using 12
spline functions. Evaluation metrics, aggregated over the 10 time points,
can be found in Figure~\ref{fig:sim_block}.

Comparing \texttt{LSVCMM} to its independent counterpart, \texttt{LSVCM},
we find that the largest difference appears in the estimation error, especially
for weak signal (large variance or large random effect). The cross-sectional
method, \texttt{ALasso}, generally does worse in estimation and accuracy, except 
in very strong signal regimes, where it outperforms all others in accuracy. 
\texttt{SPFDA} performs similarly to \texttt{LSVCMM}
in terms of estimation, apart from strong signal regimes, but does much worse in terms
of support recovery. Additional metrics, featured in Figure~S1 of the 
Supplementary Material show that \texttt{SPFDA} typically selects more
time points as differential, leading to largely inflated FDR and slightly better power.
In particular, as the proportion of missing data decreases, we would expect
\texttt{SPFDA} to become comparable to \texttt{LSVCMM}, but accuracy actually decreases,
interestingly.

\subsection{Irregular Sampling}

\begin{figure}
    \centering
    \includegraphics[width=\linewidth]{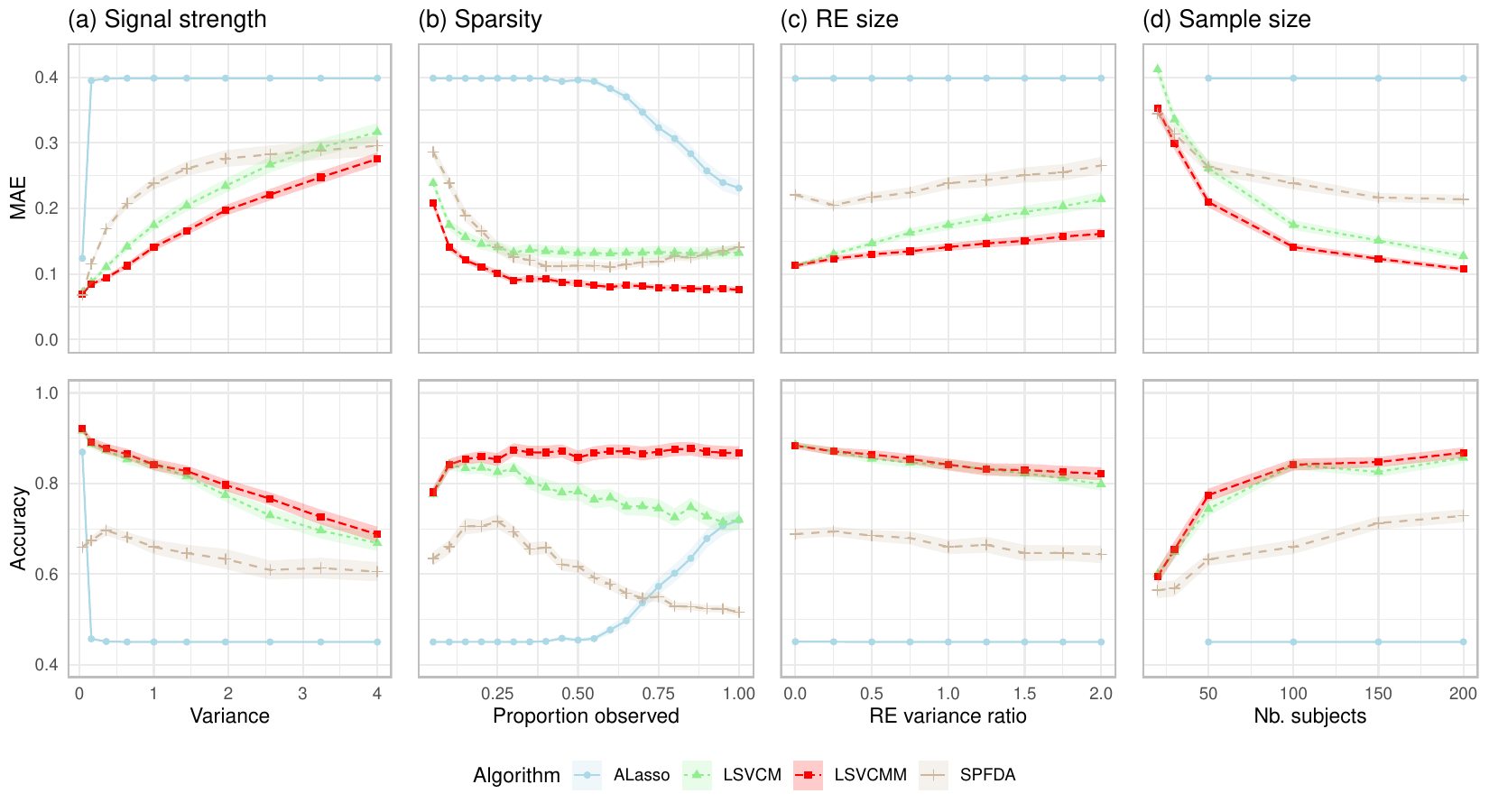}
    \caption{Evaluation metrics in the irregular sampling scenario reported as the mean
    (line) and standard error (band) across 100 replications.}
    \label{fig:sim_sparse}
\end{figure}

In this scenario, we mimic uniform sampling by using 100 regularly-spaced
time points over the unit interval and uniformly draw 10 of those as observed for 
each subject. Then, each of the 100 time points is observed, on average, only
10 times. The intercept function is again chosen to be the zero function and the
group difference function as
$\text{sigmoid}(20(0.6-t))\ind[t<0.45]$, such that the first 45 time points are null.
\texttt{SPFDA} is fitted using 50 splines. Evaluation metrics, aggregated over the 100 time points,
can be found in Figure~\ref{fig:sim_sparse}.

Cross-sectional methods have a much harder time with this setting since
there are only a few observations at each time points: indeed, \texttt{ALasso}
requires over 50 samples per time point before starting to estimate non-zero differences.
In terms of dropping longitudinal effects (\texttt{LSVCM}), we observe the same pattern
as in the missing data scenario, where weak signal and strong dependency produce worse performance. Additionally, 
we notice a much larger drop in accuracy as the sampling time become dense, where
longitudinal effects are more noticeable. This setting is also more difficult for 
\texttt{SPFDA} as it will rely more heavily on the imputed observations: the
transition from sparsely-observed data to densely-observed data makes this abundantly clear
as the estimation error becomes comparable. \add{Additionally, increasing the sample size widens the gap between \texttt{SPFDA} and \texttt{LSVCMM}, indicating that the imputation bias is a key component in explaining the difference in performance.}

\section{Oral Cancer Development Mouse Study}
\label{sec:real}

\add{The tumor suppressor gene Dmbt1 (\textit{deleted in malignant brain tumors 1}) plays a critical role in the progression of oral squamous cell carcinoma (SCC) \citep{singh_squamous_2021}. Dmbt1 is present in human saliva, where it exhibits antimicrobial properties \citep{reichhardt_salsa-dance_2017}. In a recent longitudinal study, \blindedinline{\cite{medeiros_salivary_2023}} found that Dmbt1 levels varied with cancer progression and treatment and were associated with changes in microbial composition. To better understand the interaction between Dmbt1, microbial composition, and oral SCC
development, a new mouse study was conducted \blindedinline{(Medeiros et al, manuscript in preparation)} were 76 mice were
bred with (\textit{wild type}, WT) and without (\textit{knockout}, KO) the Dmbt1 gene before being
inoculated with oral SCC. The saliva samples were then collected over time (0, 4, 8, 12, 16 and 22 weeks
after inoculation) and 16S sequencing was performed (16S rRNA, 97\% sequence similarity OTU binning). Finally, the histopathology of the tongue was evaluated at week 22 where the mice were diagnosed with pre-cancer 
\textit{epithelial dysplasia} (ED) or 
\textit{carcinoma in situ} (CIS), or with SCC.
The results strengthen the original findings of \cite{singh_squamous_2021}
as 17 of the 34 (50\%) knockout mice and only 6 of the 42 (14\%) wild type mice developed SCC by week 22, suggesting a
\textit{causal} link between Dmbt1 and cancer progression.}

\add{A potential avenue of action of Dmbt1 on cancer progression, as suggested by the
findings in \blindedinline{\cite{medeiros_salivary_2023}}, is through the microbiota.
Therefore, one of the objectives of the study is to
investigate the longitudinal association of microbial composition
with the Dmbt1 genotype (WT vs. KO) and with diagnosis (dichotomized as precancer, ED / CIS, vs. cancer, SCC) \drop{as well as the
interaction between genotype and diagnosis}. In particular, the identification of specific OTUs and weeks
with differential abundance between any of the subgroups can reveal how Dmbt1 and the microbial composition
influence cancer progression, potentially leading to better prediction of treatment response
and individualized treatments.
To this end, we consider the following VCM for the
\textit{centered log-ratio} (clr) transformed abundance of each OTU:
\begin{align}
    \bbE\{\clr(y(w))\mid \bx\}
    &=
    \beta_0(w) 
    + \beta_{\text{KO}}(w)x_{\text{KO}}\nonumber
    + \beta_{\text{SCC}}(w)x_{\text{SCC}} \\
    &\qquad+ \beta_{\text{KO:SCC}}(w)x_{\text{KO}}x_{\text{SCC}}
    + \beta_{\text{F}}(w)x_{\text{F}},
    \label{eq:introduction.vcm}
\end{align}
where $w$ denotes the week and where $x_{\text{KO}}, x_{\text{SCC}}$ and $x_{\text{F}}$
are group indicators for genotype, diagnosis and sex, respectively.}

\begin{figure}
    \centering
    \includegraphics[width=\linewidth]{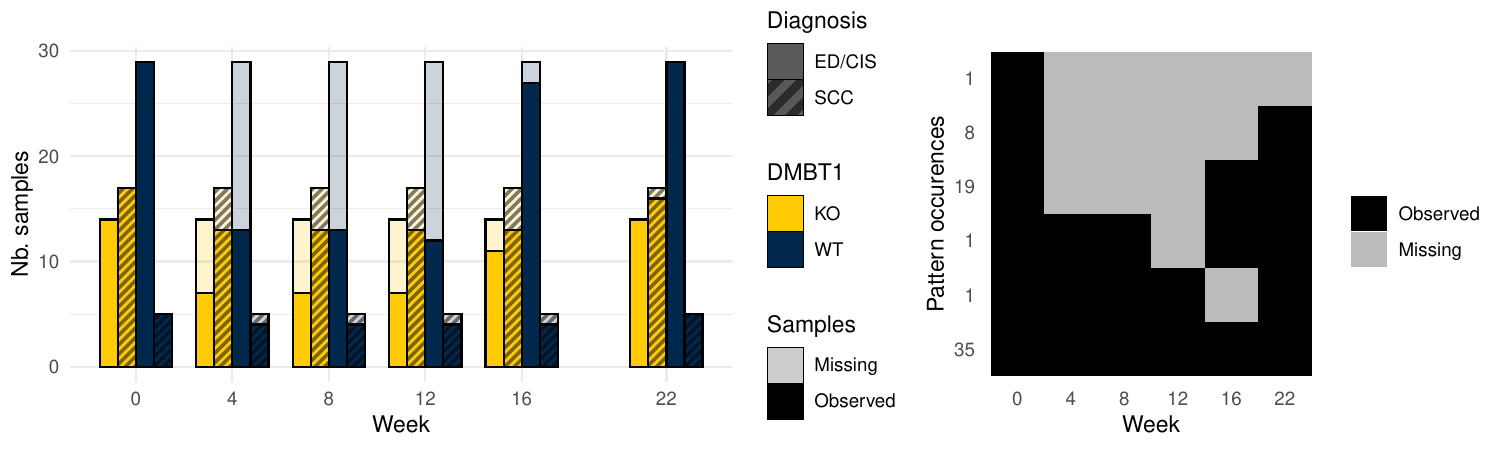}
    \caption{(Left) Number of observed and missing samples per week, stratified by genotype and diagnosis.
    (Right) Patterns of missingness; row labels indicate the frequency of that pattern occurring among the 65 mice.}
    \label{fig:dmbt.missingness}
\end{figure}

\add{An important challenge that emerges from data collection is a significant amount of missing data. Saliva samples were collected for only 65 mice and, of the $6\times 65=390$ potential samples, only 294 (75\%) were ultimately sequenced. 
Figure~\ref{fig:dmbt.missingness} shows the number of samples available per week and per subgroup, along the missingness patterns. }

We apply the \texttt{LSVCMM} methodology to VCM \eqref{eq:introduction.vcm}. After filtering out OTUs below 5\% prevalence across the 294 samples, there remains 187 OTUs to be used as the response.
We fit \texttt{LSVCMM} with a compound symmetry working covariance, with a Gaussian kernel,
with a mixed penalty ($\alpha=0.5$) to encourage global sparsity in each of the terms as well
as local sparsity to identify weeks of differential abundance. The intercept varying coefficient
is not penalized since there is no expectation it should be close to 0.
The regularization parameter $\lambda$ and the kernel scale $h$ is selected using the EBIC proposed in
Section~S1.3 of the Supplementary Material. We compare \texttt{LSVCMM} to \texttt{ALasso} 
(i.e., cross-sectional with sparsity),  and to \texttt{SPFDA} 
(with fPCA imputation of the missing entires). 
We omit \texttt{LSVCM} from the comparison since, as was seen from the 
simulation studies in Section~\ref{sec:simulations}, the main difference is in estimation error,
not selection accuracy.
For \texttt{LSVCMM} and \texttt{ALasso}, we use bootstrap to obtain simultaneous confidence bands
for all varying coefficients of interests. \texttt{SPFDA}
only provides point-wise standard errors: we produce simultaneous bands using the 
provided standard error and a Bonferroni adjustment, corresponding to multiplying the standard
error by a factor of $2.64$, the upper $1-0.025/6$ upper quantile of a standard normal distribution. 
For each of the
two main effects of genotype and diagnosis and for their interaction terms, we report the 
estimated varying coefficients and note which weeks are such that zero was excluded. For brevity,
only taxa identified as DA for at least one week by one method are reported, though 187 OTUs 
were processed.

\begin{figure}
    \centering
    \includegraphics[width=0.8\linewidth]{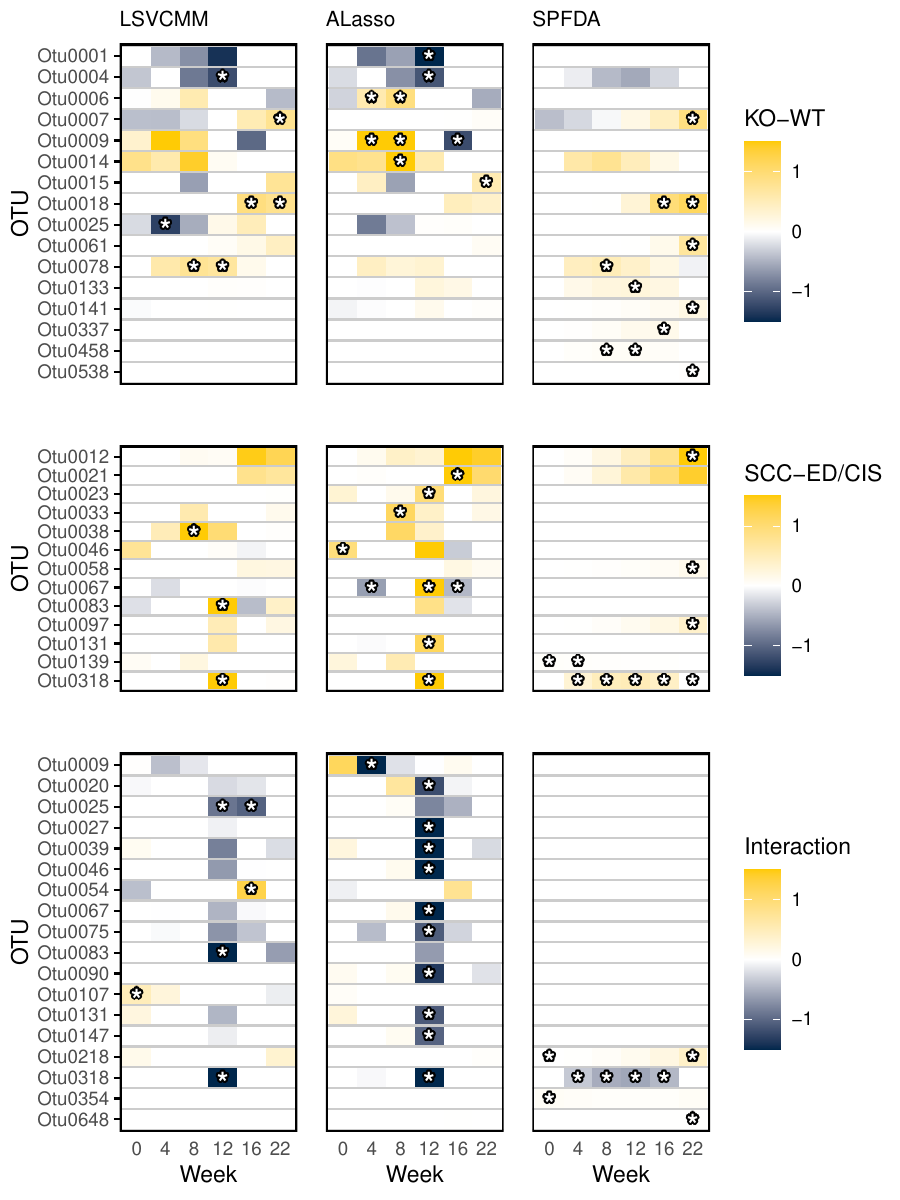}
    \caption{Estimated main effects (top: KO less WT; middle: SCC less ED/CIS) and interaction (bottom: coded as 1 if KO and SCC, and 0 otherwise). 
    White cells correspond to a zero estimate, colored cells correspond to a non-zero estimate.
    Asterisks indicate a time point where the 95\% simultaneous confidence band excludes zero.
    Columns represent estimate emerging from three different methods. Only OTUs with a significant difference
    for at least one time point and one method are included in each row (out of 187 total OTUs).}
    \label{fig:dmbt1}
\end{figure}

Figure~\ref{fig:dmbt1} contains the estimated varying coefficients for the main
effects of genotype and diagnosis as well as for the interaction term. 
We generally find strong agreement between \texttt{LSVCMM} and its cross-sectional 
counter-part \texttt{ALasso} and some agreement between \texttt{LSVCMM} and \texttt{SPFDA}. 
\texttt{ALasso} tends to select more differences than \texttt{LSVCMM}, but many of the
additional discoveries occur for the weeks 4-16 where there are significantly more missing
data. This is particularly obvious for the interaction term, where \texttt{ALasso} selects 
many OTUs at week 12. 
We note that \texttt{SPFDA} finds multiple small group differences among rare taxa
(large OTU number) which are not corroborated by \texttt{LSVCMM} nor \texttt{ALasso}, 
suggesting that the standard error estimation may be inappropriate for those instances.

Five OTUs are identified by \texttt{LSVCMM} as having a non-zero interaction term,
suggesting some interplay between the Dmbt1 gene, the abundance of those OTUs and cancer
development. In particular, OTU 0107 
identifies a positive
difference in the interaction term at week 0 and no differences in the main effects, suggesting that the
KO mice with initially higher abundance of that OTU were more likely to develop SCC by week 22.
Three OTUs, 0025, 0083 and 0318, have a negative interaction estimate at week 12. 
OTU 0025 
does not have a non-zero main effect estimate, suggesting that KO mice with
lower abundance of that OTU between week 12 and 16 were more likely to develop SCC by week 22.
OTUs 0083 and 0318 
also show a positive difference between diagnosis, indicating 
that WT mice with higher abundance of those two OTUs at week 12 were more likely to be diagnosed with SCC
by week 22. Finally, OTU 0054 
has a positive interaction estimate at week 16 and no non-zero
main effect estimates, suggesting that KO mice with higher abundance of that OTU at week 16
were more likely to be diagnosed with SCC.
All identified OTUs belong to taxonomic families known to be associated with 
oral SCC \citep{ahn_periodontal_2012, olsen_possible_2019} or with other digestive tract cancers
\citep{flemer_oral_2018, dong_detection_2019, yoon_bifidobacterium_2021}.

\section{Discussion}
\label{sec:discussion}

\texttt{LSVCMM} addresses two main deficiencies in existing sparse VCM methodologies. First, 
it includes longitudinal effects in the form of a parametric working covariance model,
whereas many methods cannot account for within-subject dependencies 
\citep{wang_functional_2015, kong_domain_2015, tu_estimation_2020, zhong_locally_2022}.
Simulation experiments in Section~\ref{sec:simulations} have shown that omitting the dependency
leads to worse estimation accuracy and worse support recovery in the densely sampled
regime. Second, our proposed approach applies to any sampling design: in particular, 
it allows missing data in regular designs as well as irregular designs, where the only method
including longitudinal effects \citep[\texttt{SPFDA}]{wang_functional_2022} is limited to regular designs.
Experiments show that imputation is insufficient in extending \texttt{SPFDA} to irregular cases,
as support recovery is inferior to \texttt{LSVCMM}.

There are multiple avenues of improvement, not only in terms of general enhancements, but also with respect to the longitudinal differential abundance application of Section~\ref{sec:real}.
First, the bootstrap simultaneous confidence bands are expensive to compute: indeed, in order
to have a granular enough estimate of the quantiles, thousands of samples are required. 
A natural alternative would be to consider ideas from the \textit{de-biased Lasso} literature
\citep{zhang_confidence_2014, honda_-biased_2021}. 
Second, while computationally convenient, the
working covariance model requires some assumption on the dependency structure which could
lead to misspecification. In Section~S2.2 of the Supplementary Materials, we consider an experiment
where \replace{the true generating covariance is AR(1), while a compound symmetry structure is fitted. We found that \texttt{LSVCMM} still improves on the independent model, even when misspecified.}{the working covariance is misspecified and found that \texttt{LSVCMM} still improves on the independent model and on \texttt{SPFDA}.}
That being said, more severe misspecification could be troublesome and an alternative treatment
of the dependency might be more robust. 
Third, the locally-constant kernel smoothing was similarly chosen for the ease of calculation, 
but it introduces some bias at the boundary of the domain and in regions or sharp variations: 
another important extension involves higher-order approximations such as the commonly-used
local linear approximation.
Fourth, utilizing a least square objective for the log-transformed
microbial abundances is generally inappropriate, largely due to the zero-inflation occurring 
for rarer taxa. For example, 78 of the 187 OTUs considered have more than 90\% of zeros
across the 265 samples and 173 have more than 50\% of zeros. 
A natural extension of \texttt{LSVCMM} would be to allow more general
distribution, similarly to GLMs; in particular, a negative binomial or Tweedie compound Poisson-Gamma objective would be of
interest for our application. Of note, \cite{zhong_locally_2022} is defined for 
\textit{generalized} VCM, though it lacks longitudinal effects. 
Fifth, we processed each OTU independently, and there are two important
sources of dependency among them that are thus disregarded. The sequencing procedure introduces 
negative dependence between OTUs because of compositionality effects, and OTUs corresponding to
related species may have positive dependency which could be captured by the taxonomic tree. 
Both of these suggest a multivariate extension.

\backmatter

\section*{Acknowledgements}

Nisha J D'Silva's research was supported by the National Institutes of Health grant R35DE027551.
Gen Li's research was partially supported by the National Institutes of Health grant R03DE031296.

This research was supported in part through computational resources and services provided by Advanced Research Computing at the University of Michigan, Ann Arbor. Computations for this research were also performed in part on the Pennsylvania State University’s Institute for Computational and Data Sciences’ Roar Collab supercomputer.

\vspace*{-8pt}

\section*{Supplementary Materials}

\begin{description}
\item[Appendix:] Additional details on estimation (S1.1 and S1.2), tuning parameter
selection (S1.3), simultaneous confidence bands (S1.4) and significance (S1.5), additional simulations results (S2.1), 
and misspecification experiments (S2.2). (PDF file)

\item[R package for LSVCMM:] R-package \texttt{LSVCMM} implementing
the proposed methodology. (GNU zipped tar file; see also the GitHub repository \blindedinline{\href{https://github.com/fontaine618/LSVCMM}{github.com/fontaine618/LSVCMM}}, version 0.0.5)

\item[R scripts:] Code for running the experiments and producing the results included in the text. (GNU zipped tar file; see also the GitHub repository \blindedinline{\href{https://github.com/fontaine618/LSVCMM-Experiments}{github.com/fontaine618/LSVCMM-Experiments}})

\end{description}\vspace*{-8pt}

\section*{Data Availability}

The data underlying this article will be shared on reasonable request to the corresponding author.

\vspace*{-8pt}

\bibliographystyle{biom} 
\bibliography{references.bib}

@book{gu_smoothing_2013,
	address = {New York, NY},
	series = {Springer {Series} in {Statistics}},
	title = {Smoothing {Spline} {ANOVA} {Models}},
	volume = {297},
	copyright = {https://www.springernature.com/gp/researchers/text-and-data-mining},
	isbn = {978-1-4614-5368-0 978-1-4614-5369-7},
	url = {https://link.springer.com/10.1007/978-1-4614-5369-7},
	language = {en},
	urldate = {2025-05-12},
	publisher = {Springer},
	author = {Gu, Chong},
	year = {2013},
	doi = {10.1007/978-1-4614-5369-7},
}

@article{parikh_proximal_2014,
	title = {Proximal algorithms},
	volume = {1},
	number = {3},
	journal = {Foundations and trends® in Optimization},
	author = {Parikh, Neal and Boyd, Stephen and {others}},
	year = {2014},
	note = {Publisher: Now Publishers, Inc.},
	pages = {127--239},
}

@article{ligtenberg_human_2001,
	title = {Human salivary agglutinin binds to lung surfactant protein-{D} and is identical with scavenger receptor protein gp-340},
	volume = {359},
	issn = {0264-6021},
	doi = {10.1042/0264-6021:3590243},
	language = {eng},
	number = {Pt 1},
	journal = {The Biochemical Journal},
	author = {Ligtenberg, T. J. and Bikker, F. J. and Groenink, J. and Tornoe, I. and Leth-Larsen, R. and Veerman, E. C. and Nieuw Amerongen, A. V. and Holmskov, U.},
	month = oct,
	year = {2001},
	pmid = {11563989},
	pmcid = {PMC1222141},
	pages = {243--248},
}

@article{reichhardt_salsa-dance_2017,
	title = {{SALSA}-{A} dance on a slippery floor with changing partners},
	volume = {89},
	issn = {1872-9142},
	doi = {10.1016/j.molimm.2017.05.029},
	language = {eng},
	journal = {Molecular Immunology},
	author = {Reichhardt, M. P. and Holmskov, U. and Meri, S.},
	month = sep,
	year = {2017},
	pmid = {28668353},
	pages = {100--110},
}

@article{singh_squamous_2021,
	title = {Squamous cell carcinoma subverts adjacent histologically normal epithelium to promote lateral invasion},
	volume = {218},
	issn = {1540-9538},
	doi = {10.1084/jem.20200944},
	language = {eng},
	number = {6},
	journal = {The Journal of Experimental Medicine},
	author = {Singh, Priyanka and Banerjee, Rajat and Piao, Songlin and Costa de Medeiros, Marcell and Bellile, Emily and Liu, Min and Damodaran Puthiya Veettil, Dilna and Schmitd, Ligia B. and Russo, Nickole and Danella, Erika and Inglehart, Ronald C. and Pineault, Kyriel M. and Wellik, Deneen M. and Wolf, Greg and D'Silva, Nisha J.},
	month = jun,
	year = {2021},
	pmid = {33835136},
	pmcid = {PMC8042603},
	pages = {e20200944},
}

@article{yoon_bifidobacterium_2021,
	title = {Bifidobacterium {Strain}-{Specific} {Enhances} the {Efficacy} of {Cancer} {Therapeutics} in {Tumor}-{Bearing} {Mice}},
	volume = {13},
	copyright = {http://creativecommons.org/licenses/by/3.0/},
	issn = {2072-6694},
	url = {https://www.mdpi.com/2072-6694/13/5/957},
	doi = {10.3390/cancers13050957},
	language = {en},
	number = {5},
	urldate = {2024-01-10},
	journal = {Cancers},
	author = {Yoon, Youngmin and Kim, Gihyeon and Jeon, Bu-Nam and Fang, Sungsoon and Park, Hansoo},
	month = jan,
	year = {2021},
	pages = {957},
}

@article{ogunrinola_human_2020,
	title = {The {Human} {Microbiome} and {Its} {Impacts} on {Health}},
	volume = {2020},
	issn = {1687-918X},
	url = {https://www.hindawi.com/journals/ijmicro/2020/8045646/?back=https%3A%2F%2Fwww.google.com%2Fsearch%3Fclient%3Dsafari%26as_qdr%3Dall%26as_occt%3Dany%26safe%3Dactive%26as_q%3DWhat%20is%20international%20microbiota%26channel%3Daplab%26source%3Da-app1%26hl%3Den},
	doi = {10.1155/2020/8045646},
	language = {en},
	urldate = {2024-01-27},
	journal = {International Journal of Microbiology},
	author = {Ogunrinola, Grace A. and Oyewale, John O. and Oshamika, Oyewumi O. and Olasehinde, Grace I.},
	month = jun,
	year = {2020},
	pages = {e8045646},
}

@article{medeiros_salivary_2023,
	title = {Salivary microbiome changes distinguish response to chemoradiotherapy in patients with oral cancer},
	volume = {11},
	copyright = {2023 The Author(s)},
	issn = {2049-2618},
	url = {https://microbiomejournal.biomedcentral.com/articles/10.1186/s40168-023-01677-w},
	doi = {10.1186/s40168-023-01677-w},
	language = {en},
	number = {1},
	urldate = {2024-01-02},
	journal = {Microbiome},
	author = {Medeiros, Marcell Costa de and The, Stephanie and Bellile, Emily and Russo, Nickole and Schmitd, Ligia and Danella, Erika and Singh, Priyanka and Banerjee, Rajat and Bassis, Christine and Murphy, George R. and Sartor, Maureen A. and Lombaert, Isabelle and Schmidt, Thomas M. and Eisbruch, Avi and Murdoch-Kinch, Carol Anne and Rozek, Laura and Wolf, Gregory T. and Li, Gen and Chen, Grace Y. and D’Silva, Nisha J.},
	month = dec,
	year = {2023},
	pages = {1--23},
}

@article{flemer_oral_2018,
	title = {The oral microbiota in colorectal cancer is distinctive and predictive},
	volume = {67},
	copyright = {© Article author(s) (or their employer(s) unless otherwise stated in the text of the article) 2018. All rights reserved. No commercial use is permitted unless otherwise expressly granted.. This is an open access article distributed in accordance with the Creative Commons Attribution Non Commercial (CC BY-NC 4.0) license, which permits others to distribute, remix, adapt, build upon this work non-commercially, and license their derivative works on different terms, provided the original work is properly cited and the use is non-commercial. See: http://creativecommons.org/licenses/by-nc/4.0/},
	issn = {0017-5749, 1468-3288},
	url = {https://gut.bmj.com/content/67/8/1454},
	doi = {10.1136/gutjnl-2017-314814},
	language = {en},
	number = {8},
	urldate = {2024-01-09},
	journal = {Gut},
	author = {Flemer, Burkhardt and Warren, Ryan D. and Barrett, Maurice P. and Cisek, Katryna and Das, Anubhav and Jeffery, Ian B. and Hurley, Eimear and O‘Riordain, Micheal and Shanahan, Fergus and O‘Toole, Paul W.},
	month = aug,
	year = {2018},
	pages = {1454--1463},
}

@article{gomaa_human_2020,
	title = {Human gut microbiota/microbiome in health and diseases: a review},
	volume = {113},
	issn = {1572-9699},
	shorttitle = {Human gut microbiota/microbiome in health and diseases},
	url = {https://doi.org/10.1007/s10482-020-01474-7},
	doi = {10.1007/s10482-020-01474-7},
	language = {en},
	number = {12},
	urldate = {2024-01-27},
	journal = {Antonie van Leeuwenhoek},
	author = {Gomaa, Eman Zakaria},
	month = dec,
	year = {2020},
	pages = {2019--2040},
}

@misc{jeganathan_block_2018,
	title = {The {Block} {Bootstrap} {Method} for {Longitudinal} {Microbiome} {Data}},
	url = {http://arxiv.org/abs/1809.01832},
	doi = {10.48550/arXiv.1809.01832},
	urldate = {2024-01-26},
	publisher = {arXiv},
	author = {Jeganathan, Pratheepa and Callahan, Benjamin J. and Proctor, Diana M. and Relman, David A. and Holmes, Susan P.},
	month = nov,
	year = {2018},
	note = {arXiv:1809.01832 [stat]},
}

@article{chen_extended_2008,
	title = {Extended {Bayesian} information criteria for model selection with large model spaces},
	volume = {95},
	issn = {0006-3444},
	url = {https://doi.org/10.1093/biomet/asn034},
	doi = {10.1093/biomet/asn034},
	number = {3},
	urldate = {2024-01-24},
	journal = {Biometrika},
	author = {Chen, Jiahua and Chen, Zehua},
	month = sep,
	year = {2008},
	pages = {759--771},
}

@article{zhang_confidence_2014,
	title = {Confidence {Intervals} for {Low} {Dimensional} {Parameters} in {High} {Dimensional} {Linear} {Models}},
	volume = {76},
	issn = {1369-7412},
	url = {https://doi.org/10.1111/rssb.12026},
	doi = {10.1111/rssb.12026},
	number = {1},
	urldate = {2024-01-10},
	journal = {Journal of the Royal Statistical Society Series B: Statistical Methodology},
	author = {Zhang, Cun-Hui and Zhang, Stephanie S.},
	month = jan,
	year = {2014},
	pages = {217--242},
}

@article{honda_-biased_2021,
	title = {The de-biased group {Lasso} estimation for varying coefficient models},
	volume = {73},
	issn = {1572-9052},
	url = {https://doi.org/10.1007/s10463-019-00740-4},
	doi = {10.1007/s10463-019-00740-4},
	language = {en},
	number = {1},
	urldate = {2024-01-10},
	journal = {Annals of the Institute of Statistical Mathematics},
	author = {Honda, Toshio},
	month = feb,
	year = {2021},
	pages = {3--29},
}

@article{dong_detection_2019,
	title = {Detection of {Microbial} {16S} {rRNA} {Gene} in the {Serum} of {Patients} {With} {Gastric} {Cancer}},
	volume = {9},
	issn = {2234-943X},
	url = {https://www.frontiersin.org/articles/10.3389/fonc.2019.00608},
	urldate = {2024-01-10},
	journal = {Frontiers in Oncology},
	author = {Dong, Zhaogang and Chen, Bin and Pan, Hongwei and Wang, Ding and Liu, Min and Yang, Yongmei and Zou, Mingjin and Yang, Junjie and Xiao, Ke and Zhao, Rui and Zheng, Xin and Zhang, Lei and Zhang, Yi},
	year = {2019},
}

@article{ahn_periodontal_2012,
	title = {Periodontal disease, {Porphyromonas} gingivalis serum antibody levels and orodigestive cancer mortality},
	volume = {33},
	issn = {0143-3334},
	url = {https://doi.org/10.1093/carcin/bgs112},
	doi = {10.1093/carcin/bgs112},
	number = {5},
	urldate = {2024-01-09},
	journal = {Carcinogenesis},
	author = {Ahn, Jiyoung and Segers, Stephanie and Hayes, Richard B.},
	month = may,
	year = {2012},
	pages = {1055--1058},
}

@article{olsen_possible_2019,
	title = {Possible role of {Porphyromonas} gingivalis in orodigestive cancers},
	volume = {11},
	issn = {2000-2297},
	url = {https://www.ncbi.nlm.nih.gov/pmc/articles/PMC6327928/},
	doi = {10.1080/20002297.2018.1563410},
	number = {1},
	urldate = {2024-01-09},
	journal = {Journal of Oral Microbiology},
	author = {Olsen, Ingar and Yilmaz, Ozlem},
	month = jan,
	year = {2019},
	pmid = {30671195},
	pmcid = {PMC6327928},
	pages = {1563410},
}

@article{fan_semiparametric_2008,
	title = {Semiparametric {Estimation} of {Covariance} {Matrixes} for {Longitudinal} {Data}},
	volume = {103},
	issn = {0162-1459},
	doi = {10.1198/016214508000000742},
	number = {484},
	urldate = {2023-06-16},
	journal = {Journal of the American Statistical Association},
	author = {Fan, Jianqing and Wu, Yichao},
	month = dec,
	year = {2008},
	pages = {1520--1533},
}

@article{goldsmith_corrected_2013,
	title = {Corrected {Confidence} {Bands} for {Functional} {Data} {Using} {Principal} {Components}},
	volume = {69},
	copyright = {Copyright © 2013, The International Biometric Society},
	issn = {1541-0420},
	url = {https://onlinelibrary.wiley.com/doi/abs/10.1111/j.1541-0420.2012.01808.x},
	doi = {10.1111/j.1541-0420.2012.01808.x},
	language = {en},
	number = {1},
	urldate = {2023-11-23},
	journal = {Biometrics},
	author = {Goldsmith, J. and Greven, S. and Crainiceanu, C.},
	year = {2013},
	pages = {41--51},
}

@article{wang_penalized_2012,
	title = {Penalized {Generalized} {Estimating} {Equations} for {High}-{Dimensional} {Longitudinal} {Data} {Analysis}},
	volume = {68},
	copyright = {© 2011, The International Biometric Society},
	issn = {1541-0420},
	doi = {10.1111/j.1541-0420.2011.01678.x},
	language = {en},
	number = {2},
	urldate = {2023-11-13},
	journal = {Biometrics},
	author = {Wang, Lan and Zhou, Jianhui and Qu, Annie},
	year = {2012},
	pages = {353--360},
}

@article{montiel_olea_simultaneous_2019,
	title = {Simultaneous confidence bands: {Theory}, implementation, and an application to {SVARs}},
	volume = {34},
	copyright = {© 2018 John Wiley \& Sons, Ltd.},
	issn = {1099-1255},
	shorttitle = {Simultaneous confidence bands},
	url = {https://onlinelibrary.wiley.com/doi/abs/10.1002/jae.2656},
	doi = {10.1002/jae.2656},
	language = {en},
	number = {1},
	urldate = {2023-11-16},
	journal = {Journal of Applied Econometrics},
	author = {Montiel Olea, José Luis and Plagborg-Møller, Mikkel},
	year = {2019},
	pages = {1--17},
}

@article{metwally_robust_2022,
	title = {Robust identification of temporal biomarkers in longitudinal omics studies},
	volume = {38},
	issn = {1367-4803},
	url = {https://doi.org/10.1093/bioinformatics/btac403},
	doi = {10.1093/bioinformatics/btac403},
	number = {15},
	urldate = {2024-01-03},
	journal = {Bioinformatics},
	author = {Metwally, Ahmed A and Zhang, Tom and Wu, Si and Kellogg, Ryan and Zhou, Wenyu and Contrepois, Kevin and Tang, Hua and Snyder, Michael},
	month = aug,
	year = {2022},
	pages = {3802--3811},
}

@article{metwally_metalonda_2018,
	title = {{MetaLonDA}: a flexible {R} package for identifying time intervals of differentially abundant features in metagenomic longitudinal studies},
	volume = {6},
	issn = {2049-2618},
	shorttitle = {{MetaLonDA}},
	url = {https://doi.org/10.1186/s40168-018-0402-y},
	doi = {10.1186/s40168-018-0402-y},
	language = {en},
	number = {1},
	urldate = {2024-01-03},
	journal = {Microbiome},
	author = {Metwally, Ahmed A. and Yang, Jie and Ascoli, Christian and Dai, Yang and Finn, Patricia W. and Perkins, David L.},
	month = feb,
	year = {2018},
	pages = {32},
}

@article{luo_informative_2017,
	title = {An informative approach on differential abundance analysis for time-course metagenomic sequencing data},
	volume = {33},
	issn = {1367-4803},
	url = {https://doi.org/10.1093/bioinformatics/btw828},
	doi = {10.1093/bioinformatics/btw828},
	number = {9},
	urldate = {2024-01-03},
	journal = {Bioinformatics},
	author = {Luo, Dan and Ziebell, Sara and An, Lingling},
	month = may,
	year = {2017},
	pages = {1286--1292},
}

@article{lin_locally_2017,
	title = {Locally {Sparse} {Estimator} for {Functional} {Linear} {Regression} {Models}},
	volume = {26},
	issn = {1061-8600},
	doi = {10.1080/10618600.2016.1195273},
	number = {2},
	urldate = {2023-05-30},
	journal = {Journal of Computational and Graphical Statistics},
	author = {Lin, Zhenhua and Cao, Jiguo and Wang, Liangliang and Wang, Haonan},
	month = apr,
	year = {2017},
	pages = {306--318},
}

@article{fan_analysis_2007,
	title = {Analysis of {Longitudinal} {Data} {With} {Semiparametric} {Estimation} of {Covariance} {Function}},
	volume = {102},
	issn = {0162-1459},
	doi = {10.1198/016214507000000095},
	number = {478},
	urldate = {2023-05-31},
	journal = {Journal of the American Statistical Association},
	author = {Fan, Jianqing and Huang, Tao and Li, Runze},
	month = jun,
	year = {2007},
	pmid = {19707537},
	pages = {632--641},
}

@article{johnson_penalized_2008,
	title = {Penalized {Estimating} {Functions} and {Variable} {Selection} in {Semiparametric} {Regression} {Models}},
	volume = {103},
	issn = {0162-1459},
	doi = {10.1198/016214508000000184},
	number = {482},
	urldate = {2023-11-13},
	journal = {Journal of the American Statistical Association},
	author = {Johnson, Brent A and Lin, D. Y and Zeng, Donglin},
	month = jun,
	year = {2008},
	pages = {672--680},
}

@article{wang_efficient_2005,
	title = {Efficient {Semiparametric} {Marginal} {Estimation} for {Longitudinal}/{Clustered} {Data}},
	volume = {100},
	issn = {0162-1459},
	url = {https://doi.org/10.1198/016214504000000629},
	doi = {10.1198/016214504000000629},
	number = {469},
	urldate = {2023-05-31},
	journal = {Journal of the American Statistical Association},
	author = {Wang, Naisyin and Carroll, Raymond J and Lin, Xihong},
	month = mar,
	year = {2005},
	pages = {147--157},
}

@article{nardi_asymptotic_2008,
	title = {On the asymptotic properties of the group lasso estimator for linear models},
	volume = {2},
	issn = {1935-7524, 1935-7524},
	url = {https://projecteuclid.org/journals/electronic-journal-of-statistics/volume-2/issue-none/On-the-asymptotic-properties-of-the-group-lasso-estimator-for/10.1214/08-EJS200.full},
	doi = {10.1214/08-EJS200},
	number = {none},
	urldate = {2023-05-30},
	journal = {Electronic Journal of Statistics},
	author = {Nardi, Yuval and Rinaldo, Alessandro},
	month = jan,
	year = {2008},
	pages = {605--633},
}

@article{poignard_asymptotic_2020,
	title = {Asymptotic theory of the adaptive {Sparse} {Group} {Lasso}},
	volume = {72},
	issn = {1572-9052},
	url = {https://doi.org/10.1007/s10463-018-0692-7},
	doi = {10.1007/s10463-018-0692-7},
	language = {en},
	number = {1},
	urldate = {2023-05-30},
	journal = {Annals of the Institute of Statistical Mathematics},
	author = {Poignard, Benjamin},
	month = feb,
	year = {2020},
	pages = {297--328},
}

@article{wang_functional_2022,
	title = {Functional group bridge for simultaneous regression and support estimation},
	issn = {1541-0420},
	url = {https://onlinelibrary.wiley.com/doi/abs/10.1111/biom.13684},
	doi = {10.1111/biom.13684},
	language = {en},
	urldate = {2023-05-08},
	journal = {Biometrics},
	author = {Wang, Zhengjia and Magnotti, John and Beauchamp, Michael S. and Li, Meng},
	year = {2022},
}

@misc{friedman_note_2010,
	title = {A note on the group lasso and a sparse group lasso},
	url = {http://arxiv.org/abs/1001.0736},
	doi = {10.48550/arXiv.1001.0736},
	urldate = {2023-05-30},
	publisher = {arXiv},
	author = {Friedman, J. and Hastie, T. and Tibshirani, R.},
	month = jan,
	year = {2010},
	note = {arXiv:1001.0736 [math, stat]},
}

@article{tibshirani_regression_1996,
	title = {Regression {Shrinkage} and {Selection} {Via} the {Lasso}},
	volume = {58},
	copyright = {© 1996 Royal Statistical Society},
	issn = {2517-6161},
	url = {https://onlinelibrary.wiley.com/doi/abs/10.1111/j.2517-6161.1996.tb02080.x},
	doi = {10.1111/j.2517-6161.1996.tb02080.x},
	language = {en},
	number = {1},
	urldate = {2023-05-30},
	journal = {Journal of the Royal Statistical Society: Series B (Methodological)},
	author = {Tibshirani, Robert},
	year = {1996},
	pages = {267--288},
}

@article{simon_sparse-group_2013,
	title = {A {Sparse}-{Group} {Lasso}},
	volume = {22},
	issn = {1061-8600},
	url = {https://doi.org/10.1080/10618600.2012.681250},
	doi = {10.1080/10618600.2012.681250},
	number = {2},
	urldate = {2023-05-30},
	journal = {Journal of Computational and Graphical Statistics},
	author = {Simon, Noah and Friedman, Jerome and Hastie, Trevor and Tibshirani, Robert},
	month = apr,
	year = {2013},
	pages = {231--245},
}

@article{zou_adaptive_2006,
	title = {The {Adaptive} {Lasso} and {Its} {Oracle} {Properties}},
	volume = {101},
	issn = {0162-1459},
	url = {https://doi.org/10.1198/016214506000000735},
	doi = {10.1198/016214506000000735},
	number = {476},
	urldate = {2023-05-30},
	journal = {Journal of the American Statistical Association},
	author = {Zou, Hui},
	month = dec,
	year = {2006},
	pages = {1418--1429},
}

@article{shields-cutler_splinectomer_2018,
	title = {{SplinectomeR} {Enables} {Group} {Comparisons} in {Longitudinal} {Microbiome} {Studies}},
	volume = {9},
	issn = {1664-302X},
	url = {https://www.frontiersin.org/articles/10.3389/fmicb.2018.00785},
	urldate = {2023-05-08},
	journal = {Frontiers in Microbiology},
	author = {Shields-Cutler, Robin R. and Al-Ghalith, Gabe A. and Yassour, Moran and Knights, Dan},
	year = {2018},
}

@article{noh_sparse_2010,
	title = {Sparse {Varying} {Coefficient} {Models} for {Longitudinal} {Data}},
	volume = {20},
	issn = {1017-0405},
	url = {https://www.jstor.org/stable/24309486},
	number = {3},
	urldate = {2023-05-08},
	journal = {Statistica Sinica},
	author = {Noh, Hoh Suk and Park, Byeong U.},
	year = {2010},
	pages = {1183--1202},
}

@misc{paulson_longitudinal_2017,
	title = {Longitudinal differential abundance analysis of microbial marker-gene surveys using smoothing splines},
	copyright = {© 2017, Posted by Cold Spring Harbor Laboratory. This pre-print is available under a Creative Commons License (Attribution 4.0 International), CC BY 4.0, as described at http://creativecommons.org/licenses/by/4.0/},
	url = {https://www.biorxiv.org/content/10.1101/099457v1},
	doi = {10.1101/099457},
	language = {en},
	urldate = {2023-05-08},
	publisher = {bioRxiv},
	author = {Paulson, Joseph N. and Talukder, Hisham and Bravo, Héctor Corrada},
	month = jan,
	year = {2017},
}

@article{staicu_significance_2015,
	title = {Significance tests for functional data with complex dependence structure},
	volume = {156},
	issn = {0378-3758},
	url = {https://www.sciencedirect.com/science/article/pii/S0378375814001566},
	doi = {10.1016/j.jspi.2014.08.006},
	language = {en},
	urldate = {2023-05-08},
	journal = {Journal of Statistical Planning and Inference},
	author = {Staicu, Ana-Maria and Lahiri, Soumen N. and Carroll, Raymond J.},
	month = jan,
	year = {2015},
	pages = {1--13},
}

@article{tu_estimation_2020,
	title = {Estimation of {Functional} {Sparsity} in {Nonparametric} {Varying} {Coefficient} {Models} for {Longitudinal} {Data} {Analysis}},
	volume = {30},
	issn = {1017-0405},
	url = {https://www.jstor.org/stable/26892791},
	number = {1},
	urldate = {2023-05-08},
	journal = {Statistica Sinica},
	author = {Tu, Catherine Y. and Park, Juhyun and Wang, Haonan},
	year = {2020},
	pages = {439--465},
}

@article{wang_shrinkage_2009,
	title = {Shrinkage {Estimation} of the {Varying} {Coefficient} {Model}},
	volume = {104},
	issn = {0162-1459},
	url = {https://doi.org/10.1198/jasa.2009.0138},
	doi = {10.1198/jasa.2009.0138},
	number = {486},
	urldate = {2023-05-08},
	journal = {Journal of the American Statistical Association},
	author = {Wang, Hansheng and Xia, Yingcun},
	month = jun,
	year = {2009},
	pages = {747--757},
}

@article{wang_variable_2008,
	title = {Variable {Selection} in {Nonparametric} {Varying}-{Coefficient} {Models} for {Analysis} of {Repeated} {Measurements}},
	volume = {103},
	issn = {0162-1459},
	url = {https://doi.org/10.1198/016214508000000788},
	doi = {10.1198/016214508000000788},
	number = {484},
	urldate = {2023-05-08},
	journal = {Journal of the American Statistical Association},
	author = {Wang, Lifeng and Li, Hongzhe and Huang, Jianhua Z.},
	month = dec,
	year = {2008},
	pmid = {20054431},
	pages = {1556--1569},
}

@article{wang_functional_2015,
	title = {Functional {Sparsity}: {Global} {Versus} {Local}},
	volume = {25},
	issn = {1017-0405},
	shorttitle = {Functional {Sparsity}},
	url = {https://www.jstor.org/stable/24721236},
	number = {4},
	urldate = {2023-05-08},
	journal = {Statistica Sinica},
	author = {Wang, Haonan and Kai, Bo},
	year = {2015},
	pages = {1337--1354},
}

@misc{zhong_locally_2022,
	title = {Locally sparse estimator of generalized varying coefficient model for asynchronous longitudinal data},
	url = {http://arxiv.org/abs/2206.04315},
	doi = {10.48550/arXiv.2206.04315},
	urldate = {2023-05-08},
	publisher = {arXiv},
	author = {Zhong, Rou and Zhang, Chunming and Zhang, Jingxiao},
	month = jun,
	year = {2022},
	note = {arXiv:2206.04315 [stat]},
}

@article{xue_variable_2012,
	title = {Variable selection in high-dimensional varying-coefficient models with global optimality},
	volume = {13},
	issn = {1532-4435},
	number = {null},
	journal = {The Journal of Machine Learning Research},
	author = {Xue, Lan and Qu, Annie},
	month = jun,
	year = {2012},
	pages = {1973--1998},
}

@article{lee_local_2016,
	title = {Local linear smoothing for sparse high dimensional varying coefficient models},
	volume = {10},
	issn = {1935-7524, 1935-7524},
	url = {https://projecteuclid.org/journals/electronic-journal-of-statistics/volume-10/issue-1/Local-linear-smoothing-for-sparse-high-dimensional-varying-coefficient-models/10.1214/16-EJS1110.full},
	doi = {10.1214/16-EJS1110},
	number = {1},
	urldate = {2023-05-08},
	journal = {Electronic Journal of Statistics},
	author = {Lee, Eun Ryung and Mammen, Enno},
	month = jan,
	year = {2016},
	pages = {855--894},
}

@article{kong_domain_2015,
	title = {Domain selection for the varying coefficient model via local polynomial regression},
	volume = {83},
	issn = {0167-9473},
	url = {https://www.sciencedirect.com/science/article/pii/S0167947314002965},
	doi = {10.1016/j.csda.2014.10.004},
	language = {en},
	urldate = {2023-05-08},
	journal = {Computational Statistics \& Data Analysis},
	author = {Kong, Dehan and Bondell, Howard D. and Wu, Yichao},
	month = mar,
	year = {2015},
	pages = {236--250},
}

@article{daye_sparse_2012,
	title = {A {Sparse} {Structured} {Shrinkage} {Estimator} for {Nonparametric} {Varying}-{Coefficient} {Model} {With} an {Application} in {Genomics}},
	volume = {21},
	issn = {1061-8600},
	url = {https://doi.org/10.1198/jcgs.2011.10102},
	doi = {10.1198/jcgs.2011.10102},
	number = {1},
	urldate = {2023-05-08},
	journal = {Journal of Computational and Graphical Statistics},
	author = {Daye, Z.   John and Xie, Jichun and Li, Hongzhe},
	month = jan,
	year = {2012},
	pmid = {22904608},
	pages = {110--133},
}

\label{lastpage}

\end{document}